\documentclass[twocolumn,final,
               prb,superbib,
               letterpaper,balancelastpage,
               lengthcheck,byrevtex,bibnotes,
               footinbib,
               ]{revtex4}
\usepackage{graphicx,amssymb,amsmath}
\usepackage{amsfonts}
\usepackage{amsmath}
\usepackage{amssymb}
\usepackage{epsf}
\usepackage{psfrag}
\usepackage{psfig}
\usepackage[pagewise,running]{lineno}
\usepackage{graphicx}
\usepackage{graphics}
\usepackage{setspace}
\usepackage{longtable}
\usepackage{lscape}
\usepackage{rotating}
\usepackage{dcolumn}
\usepackage[thinspace,squaren]{SIunits}

\def\etc{{\em etc.}}
\def\fe{{\textsc{fe}}}
\def\dd{{\mathrm{d}}}

\def\full{\protect\mbox{------}}
\def\bos{\protect\mbox{---}}
\def\kesik{\protect\mbox{--\, --\, --}}
\def\chain{\protect\mbox{-- $\cdot$ --}}

\begin{document}
\title{Numerical calculations of effective elastic properties of two cellular structures}

\author{Enis Tuncer}
\email{enis.tuncer@physics.org}

\address{Applied Condensed-Matter Physics, Department of Physics, University of Potsdam, D-14469 Potsdam Germany}

\begin{abstract}
Young's moduli of regular two-dimensional truss-like and eye-shape-like structures are simulated by using the finite element method. The structures are the idealizations of soft polymeric materials used in the electret applications. In the simulations size of the representative smallest units are varied, which changes the dimensions of the cell-walls in the structures.  A power-law expression with a quadratic as the exponential term is proposed for the effective Young's moduli of the systems as a function of the solid volume fraction. The data is divided into three regions with respect to the volume fraction; low, intermediate and high concentrations. The parameters of the proposed power-law expression in each region are later represented as a function of the structural parameters, unit-cell dimensions. The presented expression  can be used to predict structure/property relationship in materials with similar cellular structures. It is observed that the structures with volume fractions of solid higher than 0.15 exhibit the importance of the cell-wall thickness contribution in the elastic properties. The cell-wall thickness is the most significant factor to predict the effective Young's modulus of regular cellular structures at high volume fractions of solid. At lower concentrations of solid, eye-like structure yields lower Young's modulus than the truss-like structure with the similar anisotropy. Comparison of the numerical results with those of experimental data of poly(propylene) show good aggreement regarding the influence of cell-wall thickness on elastic properties of thin cellular films. 
\end{abstract}

\maketitle
\section{Introduction}
\label{sec:introducion}

In 1930's Munters and Tandberg~\cite{USPatentPS} invented the first man-made cellular materials, which opened a new era on high-tech materials. These new class of materials, called foams, were introduced in heat insulation and packaging products in early days. Nowadays, they are widely used in variety of engineering applications~\cite{CellularSolids,SuhAM2000}, such as aeronautics, impact absorbing materials~\cite{Wiklo2003,Lakes2002}, electro-mechanical sensors~\cite{ReimundRev,WegenerAPL2003}, \etc\ Recently, cellular polymers with electrical insulating properties have been shown to exhibit electro-mechanical activity\cite{ElectretPhysicsToday,Wegener-CEIDP03,ReimundRev,Reimund2000a,Reimund2000b,WegenerAPL2003,TuncerWegener2003} and are called ``ferro-electrets''.  Amplitude of stored charge inside the foam and foam's  elastic property are the main parameters that influence their electro-mechanical properties. The structure of these cellular materials is layered polymer sheet with voids (resembles a puff pastry), scanning electron micro-graphs have been  presented elsewhere\cite{ElectretPhysicsToday,WegenerAPL2003,Wegener-CEIDP03,Lekkala2001,Lekkala2000}. The materials  are produced as films with lateral dimensions much larger than the thickness one. There are possibilities to alter their internal structure with post pressure processing\cite{WegenerAPL2003,Wegener-CEIDP03} by changing the dimensions of voids. Therefore, in this paper we focus on the mechanical properties of cellular structures use in the ferro-electret applications, and assume two regular geometries, which resemble the structure of these cellular polymers.

The elastic properties in the thickness direction, effective Young's moduli, of two these structures are calculated using the finite element (\fe) method~\cite{Littmarck,femlab,Garboczi1995,Garboczi2002a,Garboczi2001Berryman} and compared with each other. The obtained effective Young's moduli of the structures are  expressed by a power-law function similar to that proposed by \citet{CellularSolids}, however, here the exponent is expressed as a quadratic function of solid volume fraction $q$, ($q=\rho^*/\rho$ with $\rho^*$ and $\rho_s$ being the densities of the cellular and solid materials). The elastic properties of the structures are expressed in the whole volume fraction range with the proposed expression as a function of structural parameters. Experimental data from literature are also presented for comparison with the numerical model.

\label{sec:background}

The cellular structure is a solid-gas mixture, a heterogeneous medium, whose macroscopic or {\em effective} properties can be expressed as a function of properties of the constituents and their volume fraction,
\begin{eqnarray}
  \label{eq:effective}
  \mathfrak{P}^{\sf e}=f(q_i,\mathfrak{P}_i)
\end{eqnarray}
where $\mathfrak{P}^{\sf e}$ is the effective property sought and subscript `i' represents the properties the constituents with volume fractions $q_i$, $\sum q_i =1$. 
For voided solids, to predict the effective Young's modulus $E^{\sf e}$ is nontrivial, due to difficulties in assigning the mechanical properties for the void. However, in such cases numerical simulations are standard methods. When the voids are open, meaning that they do not confine incompressible fluids, we can neglect this phase in the calculations. \citet{CellularSolids} have used beam theory to obtain the famous power-law relation between the effective Young's modulus $E^{\sf e}$ and the volume fraction of the solid material $q$
\begin{eqnarray}
  \label{eq:powerlaw}
  E^{\sf e}=\mathfrak{C}q^\mathfrak{n}\,E_s
\end{eqnarray}
where, $\mathfrak{C}$ and $\mathfrak{n}$ are parameters depend on the micro-structure, and $E_s$ is the Young's modulus of the solid. The value of $\mathfrak{n}$ lies between $1$ and $4$, yielding a wide range of effective properties for a given solid volume fraction $q$~\cite{Garboczi2001}. Experimental observations on various open-cell cellular materials suggest $\mathfrak{n}=2$\cite{CellularSolids}. Since the dependence of effective properties on micro-structure are not well understood in composite materials, the exact form of $\mathfrak{C}$ and $\mathfrak{n}$ are not known. Therefore to optimize and to predict foam properties numerical simulations and structure/property investigations play an important role in the current research activities~\cite{TorquatoBook,CellularSolids,TuncerWegener2003}.

\begin{figure}[t]
  \centering
    \includegraphics[]{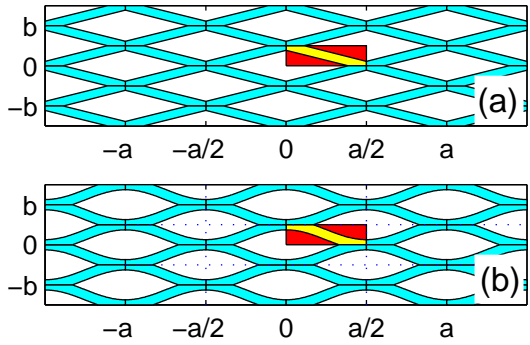}
  \caption{(a) Truss-like and (b) eye-like structures. The shaded regions are the representative units used in the calculations with the appropriate boundary conditions as shown in Fig.~\ref{fig:generalgeometry}.\label{fig:geom}}
\end{figure}

\section{Considered geometries}
\label{sec:geom-cons}


In order to simulate the elastic properties of the cellular structures in ferro-electrets presented in the literature\cite{ElectretPhysicsToday,WegenerAPL2003,Wegener-CEIDP03,Lekkala2001,Lekkala2000} two geometrical models composed  of repeating unit cells on a regular triangular lattice are constructed. In the current study influence of the structural differences are investigated. In the model geometries, the solid-void systems consist of straight and  curved layer structures as presented in Fig.~\ref{fig:geom}. The layers are connected at the  triangular lattice points. In this model, the voids become  {\em lozenge} and {\em eye-shaped} geometries, as illustrated in Fig.~\ref{fig:geom}a and ~\ref{fig:geom}b, for truss- and eye-like structures, respectively. The structures are refered as truss-like and eye-like in the text. The marked rectangular regions [$(0, 0)\le(x,y)\le(a/2, b/2)$] in Fig.~\ref{fig:geom}a and \ref{fig:geom}b are the unit cells (or the computational domain $\Omega$) used in the \fe\ calculations,
\begin{equation}
  \label{eq:3}
  \Omega=(0, 0)\le(x,y)\le(a/2, b/2)\wedge (x,y)\in\Omega_s.
\end{equation}
The parameters $a$ and $b$ are used to model different void geometries--cellular structures.  

The cell-wall thickness of the truss-like structure $t^t$ is calculated from the dimensions of the unit-cell $a$ and $b$ and volume fraction of the solid $q$.
\begin{eqnarray}
  \label{eq:thickness-truss}
  t^t(q,a,b)&=&2\sqrt{A}\,\sin (2\alpha) 
\end{eqnarray}
with
\begin{eqnarray}
  A&=&(b/2-m\tan\alpha)^2+(a/2-m)^2\nonumber\\
  m&=& \sqrt{1-q\,a\,b/(4\tan\alpha)}\nonumber\\
  \alpha&=&\arctan(b/a)
\end{eqnarray}

The cell-wall thickness of the eye-like structure is modeled with the following line expression with wall thickness $t^e$:
\begin{eqnarray}
  l_i&=&b/4[1+\cos(2\pi x/a)]+(-1)^i \cdot t^e/2 \nonumber\\
  &&\textrm{with} \quad i=\{1,2\},  \label{eq:2}
\end{eqnarray}
where $l_1$ and $l_2$ are the lower and upper lines, respectively, which represent the boundaries between the void and the cell-wall.  The region occupied by the solid, $\Omega_s(x,y) \in \Omega$, is calculated with the help of the intersection points of line $l_1$ with $y=0$, and of line $l_2$ with $y=b/2$, labeled as $x_{c_i}$, 
\begin{eqnarray}
  \label{eq:xcross}
  x_{c_i}&=&a/2 [(i-1)-1/\pi \arccos(2t^e/b-1)]\nonumber\\ 
  &&\textrm{where} \quad i=\{1,2\}
\end{eqnarray} 
The solid region is described by the discrete points $(x,y)$ that satisfy the following condition:
\begin{equation}
  \label{eq:cond1}
  \Omega_{s}=[(x,y)<l_2] \cap [(x,y)>l_1] \cap [(x,y)\in \Omega]
\end{equation}
In order to calculate the concentration (or relative density) $q$, the area of the solid region, $\Omega_{s}$, is determined numerically with an \fe\ procedure, in which the region $\Omega_{s}$ is divided into triangular subregions. The area of $\Omega_s$ is iterated with selected thicknesses $t^e$, until the desired concentration $q$ is obtained. 

\section{The numerical analysis}
\label{sec:fem}
\begin{figure}[t]
  \centering
  \psfragscanon
  \psfrag{v}{$\Omega_v$}
  \psfrag{s}{$\Omega_s$}
  \psfrag{b1}{$\partial\Omega^t$}
  \psfrag{b2}{$\partial\Omega^v$}
  \psfrag{b3}{$\partial\Omega^f$}
  \psfrag{b4}{$\partial\Omega^u$}
  \psfrag{a2}{$a$}
  \psfrag{a1}{$b$}
  \includegraphics[width=3in]{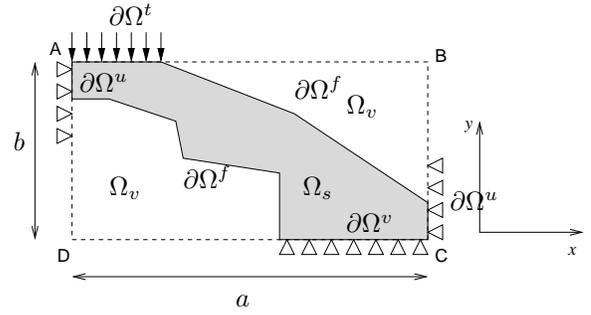}
  \psfragscanoff
  \caption{Unit cell (representative volume element) in the simulations. The region inside the dashed line (\kesik) is considered to be mirrored in all directions. The dark-gray region is the solid medium $\Omega_s$. The void is presented with $\Omega_v$. The boundary conditions are also presented as a boundary force $\partial\Omega^t$ and  constrained displacements $\partial\Omega^u$ with arrows and triangles, respectively. The free boundaries are presented as $\partial\Omega^f$. The solid region is allowed to deform only in the $y$-direction\label{fig:generalgeometry}. The dimension of the unit-cell is $a\times b$, where $a={\sf |AB|={\sf |CD|}}$ and $b={\sf |AD|}={\sf |BC|}$. }
\end{figure}

\begin{table}[t]
  \caption{Boundary conditions in the \fe. The constrained edges have $u=0$ or $v=0$ conditions for `yes' labels. All the other edges are `free' to move in corresponding directions. The applied load per unit length is presented by $F_0$.}
  \centering
  \begin{tabular*}{3.3in}{l@{\extracolsep{\fill}}rrrr}
\hline
& $\partial\Omega^t$&$\partial\Omega^u$&$\partial\Omega^v$ & $\partial\Omega^f$\\
\hline
Load & $F_0$ & --- & --- & --- \\
x-constraint & yes & yes & free & free \\
y-constraint & yes & free & yes & free \\
\hline\hline
  \end{tabular*}
  \label{tab:boundary}
\end{table}

Our \fe\ method is based on the minimization of the elastic energy in conjunction with irregular meshing~\cite{Littmarck,FEMLAB_SMM}.  The Newton's second law in the two-dimensions, which is in reality a system of two linear equations known as Navier's equations, are solved for spatial points obtained from the \fe\ triangulation of $\Omega_s$. 
\begin{eqnarray}
  \label{eq:Navier}
  -\nabla\cdot\mathbf{T}=\mathbf{K}
\end{eqnarray}
Here $\mathbf{T}$ is the mechanical stress tensor and $\mathbf{K}$ is the load vector. The stress tensor has three components in two-dimensions, which are normal stress components $\mathbf{T}_x$ and $\mathbf{T}_y$, and the shear stress component $\mathbf{S}_{xy}$,
\begin{eqnarray}
  \label{eq:Stress}
  -\frac{\partial\mathbf{T}_x}{\partial x} 
  -\frac{\partial\mathbf{S}_{xy}}{\partial y} 
  &=&\mathbf{K_x}\\
  -\frac{\partial\mathbf{T}_y}{\partial x} 
  -\frac{\partial\mathbf{S}_{xy}}{\partial y} 
&=&\mathbf{K_y}
\end{eqnarray}
Including the Hooke's law between the stress and the elastic strain, $\mathbf{T}=c\, \nabla\mathbf{u}$, lead to a  partial differential equation expressed in the global displacements $\mathbf{u}(u,v)$. Then, Eq.~(\ref{eq:Navier}) becomes as follows
\begin{eqnarray}
  \label{eq:Navier2}
  -\nabla\cdot c \nabla\mathbf{u}=\mathbf{K},
\end{eqnarray}
where the coefficient $c$ is a function of the solid materials Young's modulus and Poisson's ratio. Due to the dimensions of ferro-electret films, the contribution of the Poison-ratio is modest to elastic properties in the lateral dimensions. Our experience shows that the displacement in the thickness mode is much larger than the laterial ones. Therefore, the boundary conditions are selected such that the structures are not allowed to move in the lateral dimensions. In Fig.~\ref{fig:generalgeometry}, general geometrical considerations together with boundary conditions for the numerical model are presented. The Young's modulus of the structure is  calculated from the average deformation of the edge $\partial \Omega^t$ in the $y$-direction $\int_{\partial \Omega^t} v\,\dd x$.
\begin{eqnarray}
  \label{eq:Young}
  E^{\sf e}={b\,F_0}/{\int_{\partial \Omega^t}v\,\dd x}
\end{eqnarray}
where, $b$ is the total thickness of the unit-cell in the $y$-direction. The solid region $\Omega_s$ in Fig.~\ref{fig:generalgeometry} is used in the simulations together with the boundary conditions presented in Table~\ref{tab:boundary}. Meshing of the computational domain is an important factor in the \fe\ analysis, we have therefore adapted a procedure which used the cell-wall thickness $t$ as the meshing parameter. In the calculations, the \fe\ triangle sides are adjusted not larger than the one-tenth of the cell-wall thickness. This procedure allowed us to have small discretization units even for very thin cell-walls thicknesses efficiently. In the simulations, the volume fraction of the solid and the dimensions of the unit-cell are varied.

\section{Results and discussion}
\label{sec:results-discussion}
\subsection{Effective properties}
\begin{figure}[b]
  \centering
  \begin{tabular*}{16cm}{cc}
    \includegraphics[]{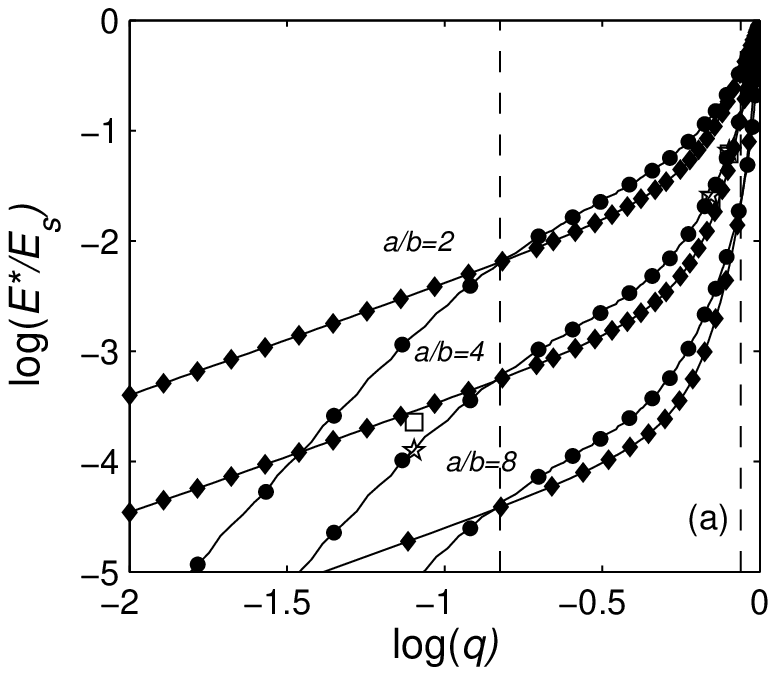}\\
    \includegraphics[]{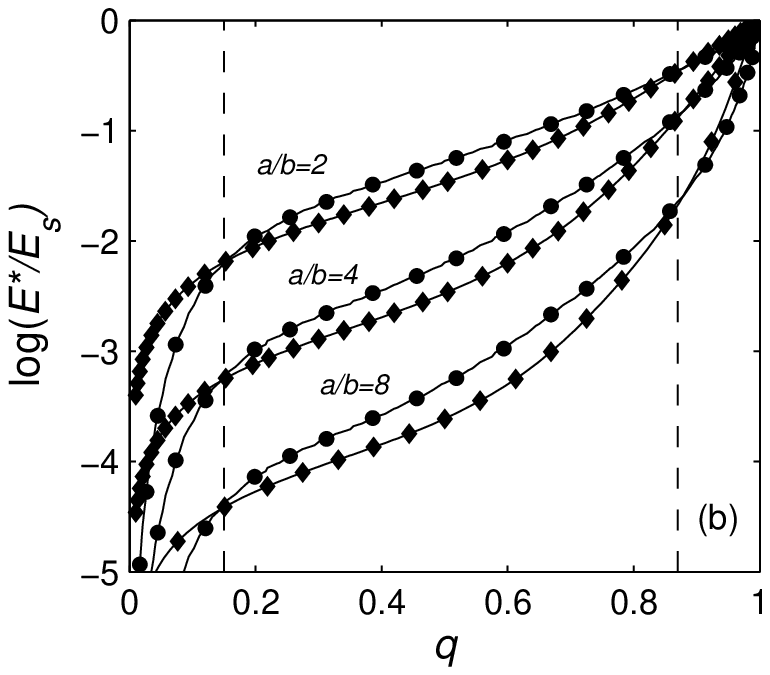}
  \end{tabular*}
  \caption{Normalized Young's moduli of truss-like ($\bullet$) and eye-like ($\blacksquare$) structures; (a) in log-log and (b) in semi-log representations. The dashed lines (\kesik) indicate the positions of volume fractions at which the elastic properties of the structures crossover each other. The solid lines (\full) are drawn to guide the eyes.\label{fig:fig248}}
\end{figure}

Simulation results for three different unit-cell dimensions are presented in Fig.~\ref{fig:fig248} as $\log$-$\log$ (\ref{fig:fig248}a) and semi-$\log$ (\ref{fig:fig248}b) plots for $a/b=\{2,\,4,\,8\}$. Two significant characteristics are observed (i) as $a/b$ increases, the normalized effective Young's moduli of the structures decrease and (b) at some volume fraction the dependence of the effective Young's moduli on $q$ becomes linear in the $\log$-$\log$ representation. In addition, when $q\approx0.15$ and $q\approx0.85$ the effective Young's moduli of the truss-like and eye-like structures crossover, which are illustrated marked with the dashed lines (\kesik) in the figure. 

The calculated data for ten different $a/b$ ratios are analyzed with a power-law expression as quadratic form in volume fraction $q$
\begin{equation}
  \label{eq:power-law}
  E^*/E_s=\mathfrak{C}\,q^{\mathfrak{a}\,q^2 +\mathfrak{b}\,q+\mathfrak{c}}
\end{equation}
where $\mathfrak{C}$, $\mathfrak{a}$, $\mathfrak{b}$ and $\mathfrak{c}$ are parameters dependent on the geometrical quantities $a/b$ and $q$. The data is divided into three volume fraction regions; low, intermediate and high, which correspond to the regions separated with the dashed lines (\kesik) in Fig.~\ref{fig:fig248}a and \ref{fig:fig248}b. The curve-fitting analyses indicated that the parameters $\mathfrak{C}$, $\mathfrak{a}$, $\mathfrak{b}$ and $\mathfrak{c}$ can be represented as a function of the geometrical parameter $a/b$. We have adopted a polynomial dependence,
\begin{eqnarray}
  \label{eq:poly}
  y=a_4x^4+a_3x^3+a_2x^2+a_1x+a_0 
\end{eqnarray}
where $y$ is anyone of the parameters $y=\{\mathfrak{C},\,\mathfrak{a},\,\mathfrak{b},\,\mathfrak{c}\}$ and $x=-\log(a/b)$. We have obtained very good agreement between the results of the polynomial analysis and the \fe\ results. The coefficients of the polynomials for each parameter in Eq.~(\ref{eq:power-law}) are presented in Table~\ref{polynomial}. 
\begin{table*}[th]
  \caption{Polynomial dependence of the fitting parameters $\log\mathfrak{C}$, $\mathfrak{a}$, $\mathfrak{b}$ and $\mathfrak{c}$ of the power-law expression in Eq.~(\ref{eq:power-law}) for three volume fraction regions. Below $x=-\log(a/b)$, and $y=a_4x^4+a_3x^3+a_2x^2+a_1x+a_0$, $y$ being one of the fitting parameters, $y=\{\mathfrak{C},\,\mathfrak{a},\,\mathfrak{b},\,\mathfrak{c}\}$.\label{polynomial}}
  \centering
  \begin{tabular*}{5.5in}{r@{~~~\extracolsep{\fill}}rrrrr@{~~~}rrrrr}
\toprule
   & \multicolumn{5}{c}{truss-like}&    \multicolumn{5}{c}{eye-like}\\
   &$a_4$ & $a_3$ & $a_2$ & $a_1$ & $a_0$ & $a_4$ & $a_3$ & $a_2$ & $a_1$ & $a_0$ \\
\colrule
    \multicolumn{11}{c}{$q\le0.15$}\\
\colrule
$\log\mathfrak{C}$ & \bos & \bos & \bos & 3.5 & $-0.45$ & \bos & \bos & \bos & 3.5 & 1.1 \\
$\mathfrak{a}$     & \bos & \bos & \bos & \bos & $\sim(-1.4)$ & \bos & \bos & $-39$ & $-51$ &11 \\
$\mathfrak{b}$     & \bos & \bos & \bos & \bos & $\sim(-0.033)$ & \bos & \bos & 21 & 27 &3.1 \\
$\mathfrak{c}$     & \bos & \bos & \bos & \bos & $\sim(1.0)$ & \bos & \bos & 1.2 & 1.4 &3.3 \\
\colrule
    \multicolumn{11}{c}{$0.15\le q\le0.85$}\\
\colrule
$\log\mathfrak{C}$ & \bos & $2.2$ & $1.5$ & $-0.053$ & $-0.023$ & \bos &$1.75$ & $0.95$ & $0.12$ & $-0.042$ \\
$\mathfrak{a}$     & \bos & 30 & 47 & 1.2 & 1.2 & \bos &11.4 & 19.2 & 0.444 &2.42 \\
$\mathfrak{b}$     & \bos & 0.17 & $-5.2$ & $-10$ & $1.5$ & \bos &$7.99$ & $8.73$ & $-5.81$ &$-0.485$ \\
$\mathfrak{c}$     & \bos & 3.08 & 4.42 & $1.04$ &$1.43$ & \bos & 1.72 & 2.2 & $-1.32$ &1.69 \\
\colrule
    \multicolumn{11}{c}{$0.15\le q\le0.85$}\\
\colrule
$\log\mathfrak{C}$ & \bos & \bos & \bos & \bos & $\sim(0.01)$ & \bos & 0.48 & 0.58 & $0.19$ & $-.019$ \\
$\mathfrak{a}$ & 3620 & 5120 & 1870 & 217 & 0.864 & \bos & $-2500$ &$- 2500$ & $-810$ &1.9\\
$\mathfrak{b}$ & $-6240$ &$-9140$ &$-3490$ &$-422$ &3.52 & \bos & $4130$ & $4240$ & $1350$& 4.75\\ 
 $\mathfrak{c}$ & $2660$ & 4000 & 1590 & 187 & 0.17 &\bos &$-1680$ & $-1750$ &$-570$ &$-1.94$\\
\botrule
\end{tabular*}
\end{table*}

\subsection{Low volume fractions of solid}

\begin{figure}[t]
  \centering
  \psfragscanon
  \psfrag{C}{$\mathfrak{C}$}
  \psfrag{a}{$\mathfrak{a}$}
  \psfrag{b}{$\mathfrak{b}$}
  \psfrag{c}{$\mathfrak{c}$}
    \includegraphics[]{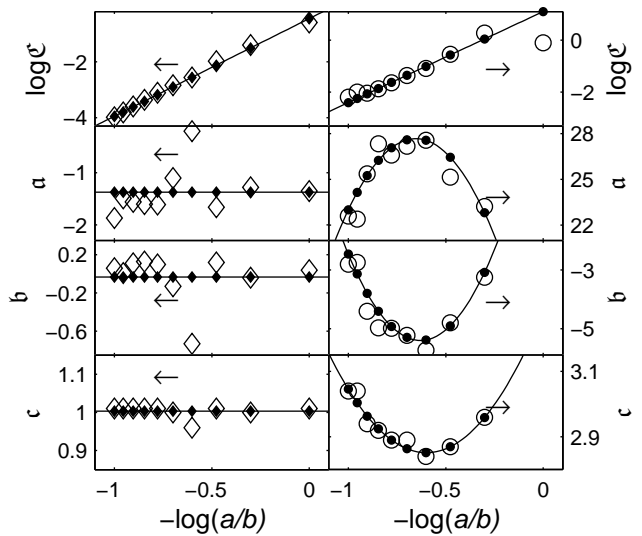}
  \psfragscanoff
  \caption{Dependence of the fitting parameters $\mathfrak{C}$, $\mathfrak{a}$, $\mathfrak{b}$ and $\mathfrak{c}$ on $a/b$ ratio. The graphs at the left and right are respectively for truss-like and eye-shape like geometries. The symbols ($\Diamond-\blacklozenge$) and ($\circ-\bullet$) represents truss-like  and eye-shaped structures respectively. The open symbols are the results of Eq.~(\ref{eq:power-law}), and the filled symbols together with the solid lines (\full) are the polynomial curve-fitting results. The parameters $\mathfrak{a}$, $\mathfrak{b}$ and $\mathfrak{c}$ are nearly constant in the considered range of $a/b$ ratios for the truss-like structure, therefore, the average values are presented in these graphs. The coeffients of the polynomials are presented in Table~\ref{polynomial}.\label{fig:dataLOW}}
\end{figure}

At low volume fractions of the solid ($q<0.15$), the truss-like structure yield higher Young's moduli than the eye-like one. The fitting parameters are presented with open symbols as a function of $a/b$ in Fig.~\ref{fig:dataLOW}. In the figure the solid lines (\full) represent the curve fitting results with the values in Table~\ref{polynomial} using Eq.~(\ref{eq:poly}). The two structures show interesting behavior when dependence of the Young's moduli are considered. The truss-like structure can be expressed by a simple relation as the one proposed by \citeauthor{CellularSolids}, Eq.~(\ref{eq:powerlaw})  with $\mathfrak{n}=\mathfrak{c}\approx1$---($\mathfrak{a}\,q^2+\mathfrak{b}\,q\ll0$ for $q \le 0.15$). In addition, there is a linear relation between pre-exponential $\log \mathfrak{C}$ in Eq.~(\ref{eq:power-law}) and $\log(a/b)$ for the truss-like structure, see Table~\ref{polynomial}. 


The eye-like structure on the other hand do not exhibit similar behavior as the truss-like structure. The pre-exponential can be expressed similar to the truss-like structure's one but the constant term is higher. The other parameters in the proposed effective formula yield quadratic relation to structure parameter $\log(a/b)$, see Table~\ref{polynomial} and Fig.~\ref{fig:dataLOW}. 
The parameters at the exponential term, $\mathfrak{a}$, $\mathfrak{b}$ and $\mathfrak{c}$, have minimum or maximum approximately around $a/b=4$ as shown in Fig.~\ref{fig:dataLOW}. Last but not least as the concentration of the solid phase is much smaller than 0.1, $\mathfrak{n}\approx\mathfrak{c}$, and equal to $1$ and $\sim3$  for truss- and eye-like structures, respectively.


\subsection{Intermediate volume fractions of solid}
\begin{figure}[t]
  \centering
  \psfragscanon
  \psfrag{C}{$\mathfrak{C}$}
  \psfrag{a}{$\mathfrak{a}$}
  \psfrag{b}{$\mathfrak{b}$}
  \psfrag{c}{$\mathfrak{c}$}
    \includegraphics[]{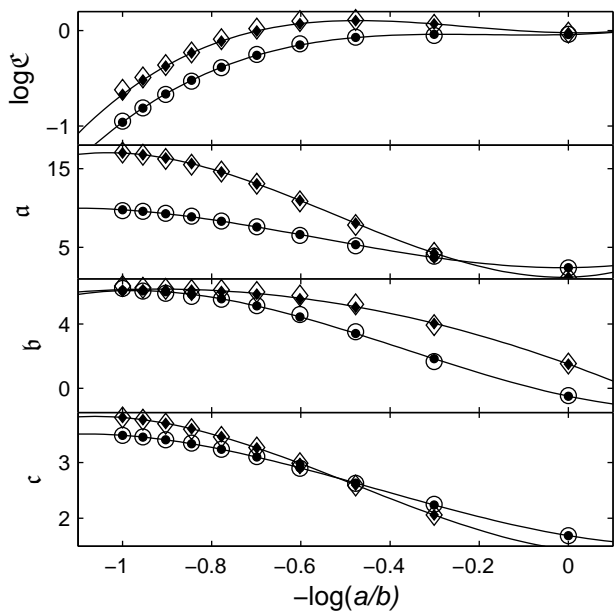}
  \psfragscanoff
  \caption{Dependence of the fitting parameters $\mathfrak{C}$, $\mathfrak{a}$, $\mathfrak{b}$ and $\mathfrak{c}$ on $a/b$ ratio. The graphs at the left and right are respectively for truss-like and eye-shape like geometries. The symbols ($\Diamond-\blacklozenge$) and ($\circ-\bullet$) represents truss-like  and eye-shaped structures respectively. The open symbols are the results of Eq.~(\ref{eq:power-law}), and the filled symbols together with the solid lines (\full) are the polynomial curve-fitting results. The coeffients of the polynomials are presented in Table~\ref{polynomial}.\label{fig:dataINT}}
\end{figure}

At intermediate volume fraction of the solid, $0.15 \le q \le 0.85$, the eye-like structure has higher elastic modulus $E^*$ than the truss-like one.  Moreover, the eye-like structure shows a moderate change with respect to $q$ compared to the truss-like one, see Fig.~\ref{fig:fig248}b. The truss-like structure indicated a bending point (a knee), while the eye-like structure does not. The fitting parameters are dependent on the unit-cell dimensions $a$ and $b$, and cubic polynomials can express the dependence as presented in Fig.~\ref{fig:dataINT} and Table~\ref{polynomial}.


\subsection{High volume fractions of solid}
\begin{figure}[t]
  \centering
  \psfragscanon
  \psfrag{C}{$\mathfrak{C}$}
  \psfrag{a}{$\mathfrak{a}$}
  \psfrag{b}{$\mathfrak{b}$}
  \psfrag{c}{$\mathfrak{c}$}
    \includegraphics[]{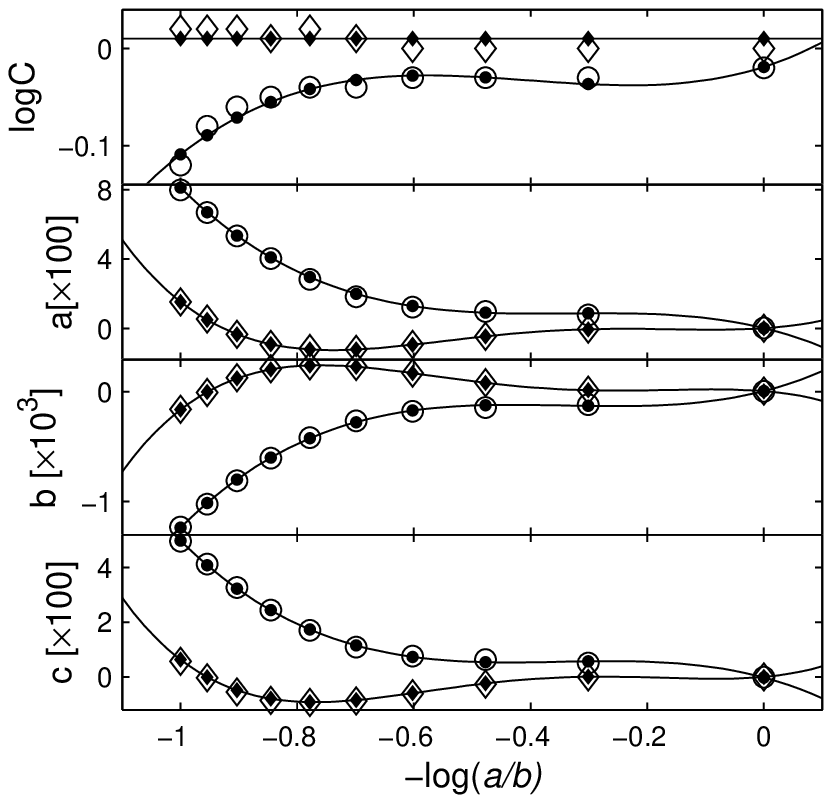}
  \psfragscanoff
  \caption{Dependence of the fitting parameters $\mathfrak{C}$, $\mathfrak{a}$, $\mathfrak{b}$ and $\mathfrak{c}$ on $a/b$ ratio. The graphs at the left and right are respectively for truss-like and eye-shape like geometries. The symbols ($\Diamond-\blacklozenge$) and ($\circ-\bullet$) represents truss-like  and eye-shaped structures respectively. The open symbols are the results of Eq.~(\ref{eq:power-law}), and the filled symbols together with the solid lines (\full) are the polynomial curve-fitting results. The parameter $\mathfrak{C}$ is nearly constant in the considered range of $a/b$ ratios for the truss-like structure, therefore, the average value is presented in the graph. The coeffients of the polynomials are presented in Table~\ref{polynomial}.\label{fig:dataHIG}}
\end{figure}

As the concentration approach to higher values $q>0.85$, the structures loose their porous character and the structural differences are not pronounceable as at the other concentration regions. In this region, the truss-like structure has higher effective Young's modulus $E^*$ than the eye-like one. The fitting parameters of the proposed effective medium equation [Eq.~(\ref{eq:power-law})] are shown in Fig.~\ref{fig:dataHIG}. The results of the dependency on $a/b$ ratio are also plotted in the figure with the polynomial coefficients presented in Table~\ref{polynomial}. One very significant difference between two structure is the form of the polynomial assumed, cubic polynomials are enough for the eye-like structure, while it is  forth order polynomials for the truss-like one in two of the coefficients. 

  

\subsection{Cell-wall thickness}
\begin{figure}[t]
  \centering
    \includegraphics[]{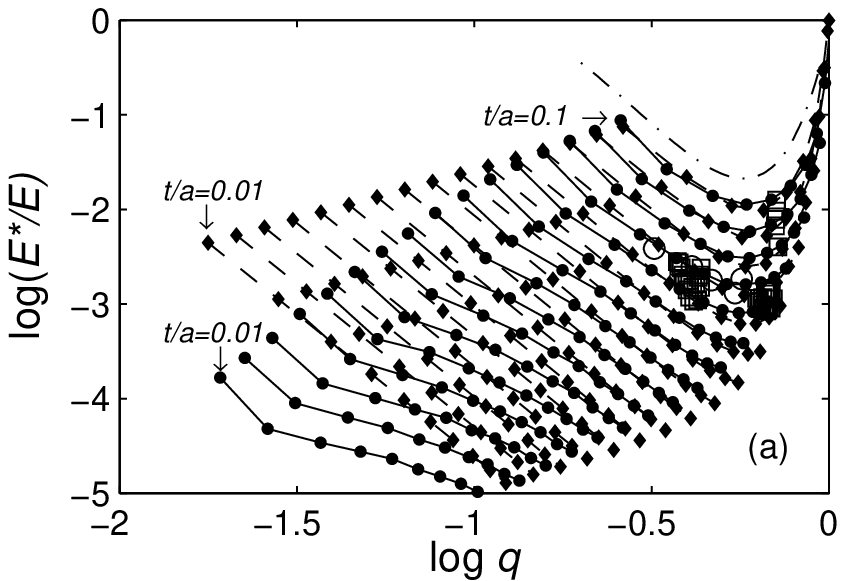}\\
    \includegraphics[]{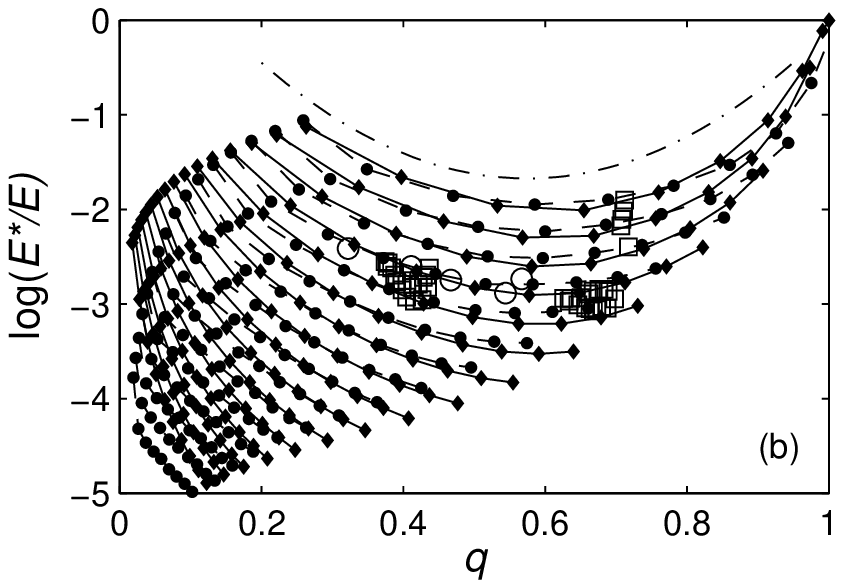}
  \caption{Normalized Young's moduli of truss-like ($\bullet$) and eye-like ($\blacksquare$) structures for various cell-wall thicknesses (a) in $\log$-$\log$ and (b) semi-$\log$ representation. The cell-wall thickness are normalized to the base of the unit-cell $a$. Sixteen thickness values are presented, $-2\le \log(t/a)\le-1$. The solid (\full) and dashed (\kesik) lines are drawn to guide the eye. The chain line (\chain) is a quadratic relation in $q$. The open symbols are the experimental data taken from the literature\cite{Wegener-CEIDP03,WegenerAPL2003}.\label{fig:thickness}}
\end{figure}

Finally, we have also investigated the influence of cell-wall thickness $t$ on the Young's modulus. In order to perform the analysis, we isolate the volume fraction $q$ in Eq.~(\ref{eq:thickness-truss}) for the thickness of truss-like structure $t^t$, and use the \fe\ modeling to calculate the thickness $t^e$ for a given volume fraction $q$ for the eye-like structure. We can  then convert Fig.~\ref{fig:fig248} to constant cell-wall thicknesses for both structures. The results are shown in Fig.~\ref{fig:thickness} for $\log$-$\log$ and semi-$\log$ representations. The values are normalized to the base length  $a$ of the unit-cells by keeping the height constant, $b=1$. It is clear that for some $a/b$ values ($4\le a/b\le 10$) and at high and intermediate concentrations of solid ($q\ge 0.15$), both structures elastic moduli yield similar values, when constant cell-wall thicknesses are considered. The chain lines (\chain) are drawn to indicate the quadratic dependence, $E^{*}(t)=b_2q^2+b_1q+b_0$. The anisotropy and structural differences do not play any significant role at high solid concentrations, and the cell-wall thickness is the main factor that influences the elastic properties. At lower concentrations of solid ($q\le 0.15$), on the other hand, the structural differences affect the Young's modulus $E^{*}$. The influence of cell-wall thickness is most significant at high $a/b$ values at low concentrations $q$. There is a minimum for the Young's modulus for a selected cell-wall thickness for both structures. The position of this minimum is around $q=0.6$, and it shows a very slight trend to move towards lower concentrations with decreasing $t/a$. It is remarkable that the constant cell-wall behavior is very distinct and we do not require to divide the $q$-axis in regions as before. 

In order to illustrate the success of the model, experimental values  from the literature \cite{Wegener-CEIDP03,WegenerAPL2003} are also plotted with open symbols in Fig.~\ref{fig:thickness}. The data were obtained by altering void dimensions of polypropylene samples by post pressure treatment.  Later the elastic proporties were measured using impedance spectroscopy\cite{AxelReview}. The density and the Young's modulus of the solid were $\rho_s=1\ \gram/\centi\meter^{3}$ and  $E_s=1\ \giga\pascal$, respectively. The experimental data display a similar behavior as predicted by the \fe\ modeling. Most of the data points sit on the equi-thickness line  $t/a\sim[0.04-0.05]$, which is in good agreement with the scanning electron micro-graphs. The proposed power-law expression in Eq.~(\ref{eq:power-law}) with the polynomial dependence on the structural parameters  can be used to prognosticate the elastic properties of cellular thin films by using their scanning electron microscope images.  
\section{Conclusions}
\label{sec:concl-disc}

In this paper, we present numerical calculations of the elastic properties of two cellular structures which resemble cellular polymers used in the electret applications. The computational domain is constructed from unit-cells (repeating units) with solid phase in the form of truss- and eye-like shapes. The size of the unit-cells are varied in the simulations to introduce anisotropy. The elastic properties of the structures are presented for various unit-cell dimensions. The structures' effective Young's moduli crossover each other at two concentration levels, $q=0.15$ and $q=0.85$. These concentration levels are used in the analysis. The most significant difference between the two structures is that the considered eye-like structure shows much lower Young's moduli than the truss-like one at low concentration of solids, which is due to the curved cell-wall. Although, at intermediate and high concentrations, there are visible differences in the Young's modulus of the structures when expressed as a function of concentration, if the cell-wall thicknesses are taken into consideration, the Young moduli exhibit similar behavior. We can therefore conclude that the cell-wall thickness is the most significant factor at high concentrations of solid in cellular materials. The influence of structural differences are only visible at low concentrations and less isotropic structures.  The calculated effective Young's moduli $E^{*}$ can be expressed with a power-law, in which the power exponent is a quadratic expression on the volume fraction of the solid $q$. The coefficients of the quadratic term and the pre-exponential are expressed as a function of the unit-cell dimensions. A full solution for the whole volume fraction spectrum is given for both structure, which can be of interest for materials researchers dealing with similar structures as presented ones. Finally, computer simulations are excellent tools to better understand the structure/property relations in cellular layered structures. 

\subsection*{Acknowledgment}
 Dr. M. Wegener is thanked for fruitful discussion.
\bibliography{../../newref.bib}
\bibliographystyle{apsrev}
\end{document}